\documentclass[11pt]{article}

\usepackage[margin=1in]{geometry}
\usepackage{authblk}
\usepackage{siunitx}
\usepackage{booktabs}
\usepackage{multirow}
\usepackage{xcolor}
\usepackage{microtype}
\usepackage{url}
\usepackage{hyperref}
\usepackage{amsmath,amssymb}
\usepackage{graphicx}
\usepackage[numbers,sort&compress]{natbib}

\title{Stable Sentiment and Persistent Dynamics in U.S. Economic News over 45 Years}

\author[1,2]{Luis E. C. Rocha}
\affil[1]{Department of Economics, Ghent University, 9000 Ghent, Belgium}
\affil[2]{Department of Physics and Astronomy, Ghent University, 9000 Ghent, Belgium}

\date{}

\begin{document}

\maketitle

\begin{abstract}
Collective emotion is often inferred from the tone of mass media, but such emotion is not directly observed. One approximation is to extract sentiment from text and use sentiment indexes as proxies to study the temporal organization of news sentiment. Using a daily index of U.S. economic news sentiment from 24 newspapers (1980--2025), we examine whether the response time of this sentiment process has changed. Although the average balance of positive and negative coverage has remained broadly stable, the persistence of news sentiment states has increased substantially. In dynamical terms, this implies longer residence times in optimistic or pessimistic regimes and weaker short-run correction of sentiment shocks. Complementary statistics show declining sentiment volatility, fewer reversals, and increasing bimodality, i.e. a stronger separation between positive and negative sentiment states. We also find an asymmetry between bursts of negative and positive sentiment, with negative bursts tending to last longer. These patterns are consistent with a minimal endogenous-memory model in which a slowly evolving latent sentiment component becomes more persistent while short-range corrective feedback weakens. The findings indicate a change in the temporal response of the U.S. economic newspaper sentiment index over the last 45 years, with sentiment shocks leaving longer traces than expected under short-memory exponential decay. News-based sentiment is thus better modeled as persistent episodes rather than as daily reactions that reset after each event.
\end{abstract}



\vspace{0.5em}
\noindent\textbf{Correspondence:} Luis E. C. Rocha, \href{mailto:luis.rocha@ugent.be}{luis.rocha@ugent.be}

\section*{Introduction}


Collective emotion, proxied by the tone of mass media and digital communication streams, is a central driver of attention and choice in complex socio-economic systems \citep{GoldenbergGross2020}. Narratives generated by news reflect and shape belief dynamics. They can become contagious stories that coordinate expectations and behavior, and sometimes detach from underlying structural drivers~\citep{Shiller2017,NimarkPitschner2019, HestonSinha2017,MaiEtAl2018}. In large-scale information ecosystems, these narratives contribute to shaping how citizens perceive society, political conflict, crises, and institutional performance. Fluctuations in news sentiment thus act as a macroscopic state variable supporting the interpretation of new events rather than merely tracking them \citep{McCombsShaw1972,Vladisavljevic2015, Verboord2023}. Text-based indicators derived from news outlets are shown to move systematically with macroeconomic conditions, policy actions, and geopolitical shocks \citep{BakerBloomDavis2016, ShapiroSudhofWilson2020}. Digital traces of collective emotion and online communication further show that responses can be long-lived, socially reinforced, and organized in clustered avalanches~\citep{RybskiEtAl2012,SanoEtAl2019,GarciaRime2019,BovetMakse2019,KarsaiEtAl2018,ZareiEtAl2024}.

The modern news ecosystem is not a passive mirror of the underlying state of the system. It is embedded in evolving media infrastructures that couple editorial decision-making, platform algorithms, and audience feedback loops~\citep{VanDijckPoellDeWaal2018,PoellNieborgVanDijck2019}. Text from news articles and social media has been used to construct measures of sentiment, attention, and uncertainty, which have been linked to macroeconomic forecasting, financial volatility, and variation in political perceptions across social groups \citep{BakerBloomDavis2016,BollenEtAl2011,PreisEtAl2013,OliveiraEtAl2017,SorokaSteculaWlezien2015,IyengarWestwood2015}. These couplings create feedback loops in which local coverage can amplify or dampen the impact of shocks on beliefs and investor trading \citep{EngelbergParsons2011}. Ranking and recommendation systems, on the other hand, reshape user exposure to diverse viewpoints and the diffusion of misinformation~\citep{BakshyMessingAdamic2015,FlaxmanGoelRao2016,VosoughiRoyAral2018, BailEtAl2018,GuessNyhanReifler2018, PerraRocha2019, DignaziEtAl2025}. Observed news sentiment can thus be understood as an emergent signal shaped by these coupled socio-technical dynamics~\citep{PetersonSalahuddinDiakopoulos2020, VanDijckPoellDeWaal2018,PoellNieborgVanDijck2019}. Automation extends these loops from curation to content production. Automated journalism systems and bots scale targeted, emotionally charged narratives~\citep{Graefe2016,Carlson2015,MontalReich2017}, while journalists and editors increasingly rely on real-time engagement metrics and social feedback when selecting and repeating stories~\citep{BlanchettNeheli2018,LamotPaulussenVanAelst2021,PetersonSalahuddinDiakopoulos2020}. Such reinforcing infrastructures suggest that changes in media organization may alter not only the level of news sentiment but also how shocks are amplified, filtered, or repeated.

Prior research has mainly used news sentiment as an indicator or predictor of economic, financial, and political outcomes. In this paper, we study the temporal memory of the sentiment process itself. We analyze both the sentiment level, to capture how long sentiment remains in a state, and its daily change, which reflects short-run correction and reversal after shocks. By using a specific lexicon-based sentiment index of economic news from major U.S.\ newspapers, we show that while the average level of the news sentiment is broadly stable over nearly five decades, its temporal organization is not. Using historical periods as reference points, we find that the social-media period is associated with stronger short-scale persistence, with sentiment states reversing less quickly than in earlier periods. We quantify long-range dependence using the Hurst exponent $H$ estimated via detrended fluctuation analysis (DFA), which captures how sentiment depends on its past across different timescales~\citep{PengEtAl1995, KantelhardtEtAl2001}. We show that the observed temporal shifts can be reproduced by a minimal endogenous-memory model in which a long-memory latent sentiment component and a short-range feedback process match the drift in Hurst exponents across media periods. The volatility and frequency of sentiment swings decline while bimodality increases, allowing strongly positive or negative sentiment states to persist. We also find an increasing asymmetry between bursts of negative and positive sentiment, with negative ones lasting longer. Our results suggest that this U.S. economic news sentiment has shifted from a more reactive process, in which shocks were corrected relatively quickly, towards a regime in which the current tone lasts longer. This implies that news sentiment should be modeled as a persistent dynamical state rather than a short-memory signal that relaxes exponentially after each event.

\subsection*{News data and sentiment index}

We analyze the daily news sentiment index introduced by Shapiro et al.~\citep{ShapiroSudhofWilson2020,data}, which was designed to measure the sentiment of economic and financial articles from U.S. newspapers. The corpus is obtained from Factiva (\url{www.dowjones.com/business-intelligence/factiva/}). To address the specificity of economic news and reduce sensitivity to lexicon choice~\citep{ChanEtAl2021}, they developed a domain-specific scoring model in two steps. First, they used the Vader sentence classifier~\citep{HuttoGilbert2014} combined with the Loughran-McDonald finance lexicon~\citep{LoughranMcDonald2011} and the Hu-Liu online reviews lexicon~\citep{HuLiu2004} to assign a positive, negative, or neutral sentiment to each sentence. Then, they counted how often each word in their full corpus appeared in each of the three classes and used pointwise mutual information to score words by their association with positive and negative sentences. The overall sentiment of a word is then the difference between the positive and negative mutual information. The final article-scoring model combines this new lexicon with the two lexicons above and applies a negation rule (i.e., multiplying the score by $-1$ if the word is preceded within three words by a negation term). The sentiment score $s_a$ of article $a$ is then the average of the word sentiment scores in $a$. A sample of 800 articles, manually rated by research assistants, indicated that this model performed better than alternative lexical models and was preferred to machine-learning methods, specifically bag-of-words, GloVe, and BERT. BERT provided similar performance in a smaller test set but was not used because of lower transparency and limited sample size.

Article-level scores are aggregated into a daily time series using a fixed-effects regression, which estimates separate effects for each day and newspaper $\times$ article-type combinations:

\begin{equation}
s_a = f_{t(a)} + f_{p(a),j(a)} + \varepsilon_a
\label{eq:01}
\end{equation}

where $t(a)$ is the article publication day, $j(a)$ is the newspaper, and $p(a)$ is the article type (editorial or regular article). The variables $f_{t(a)}$ and $f_{p(a),j(a)}$ are the sample-day and newspaper $\times$ article-type fixed effects, respectively. The day fixed effect captures the common sentiment component of all articles published on that day, while the interaction captures the usual baseline tone of each outlet/type combination. The newspaper $\times$ article-type effects remove persistent baseline differences across outlets and between editorials and regular articles, so the index is an adjusted measure of daily economic newspaper sentiment rather than a raw average of article scores. The estimated day effects ($\hat f_t=x_t$) at day $t$ give the daily sentiment index we analyze in the following sections. Values near zero should be understood as weak or balanced sentiment, not that the article or day is substantially neutral.

This method improves comparability over time, but does not fully control for changes in article volume, topic composition, journalistic style and editorial routines, semantic drift, or time-varying sentiment intensity within newspaper/type categories. At the monthly level, however, the full news sentiment index is strongly correlated ($r=0.59$) with the University of Michigan Consumer Sentiment Index (based on surveys), reaching up to $r=0.74$ after 2005~\citep{ShapiroSudhofWilson2020}. We therefore interpret the results as changes in the temporal organization of the economic-news sentiment measure, rather than as direct measurements of public emotion or the full news ecosystem.

\begin{figure}
\centering
\includegraphics[width=\columnwidth]{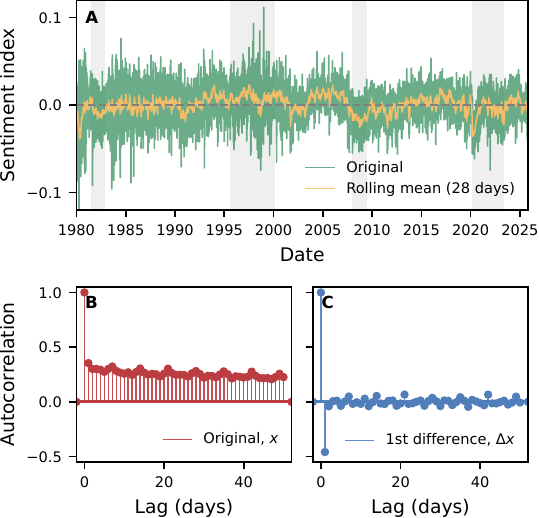}
\caption{Statistical characterization of the daily news sentiment index from 1980 to 2025. (\textbf{A}) Time series of the sentiment index $x_t$ and the rolling mean (28 days), with gray shading marking major macroeconomic events (Volcker disinflation, dot-com bubble, global financial crisis, COVID-19). Autocorrelation function (ACF) for (\textbf{B}) the sentiment index $x_t$ and its (\textbf{C}) first difference $\Delta x_t = x_t - x_{t-1}$.}
\label{fig:01}
\end{figure}

\subsection*{Temporal structure of U.S. news sentiment}

We analyze the daily index of U.S. economic news sentiment from 1~January~1980 to 27~October~2025 ($N = 16,737$ days), summarizing the aggregate economic tone across 24 U.S.-based newspapers~\citep{data}. We first characterize the basic temporal properties of the news sentiment index. Figure~\ref{fig:01}A shows that the news sentiment index $x_t$ fluctuates around a stable mean, with no evidence of large long-term drift or regime shifts in the mean. Augmented Dickey-Fuller (ADF) tests reject a unit root (-7.44, $p<0.001$), while Kwiatkowski-Phillips-Schmidt-Shin (KPSS) tests reject strict level-stationarity (0.98, $p<0.01$), consistent with a mean-reverting but persistent series. For the first differences , or daily changes, $\Delta x_t = x_t - x_{t-1}$, ADF also strongly rejects a unit root ($-31.78$, $p < 0.001$) and KPSS does not reject stationarity (0.02, $p=0.10$), indicating stationary noise around zero. These daily changes capture short-run correction and reversal. The autocorrelation function (ACF) for $x_t$ starts around 0.35 at lag 1 and decays slowly with a mild weekly oscillation, whereas the ACF of $\Delta x_t$ shows a negative spike at lag 1 and then fluctuates around zero (Figs.~\ref{fig:01}B-C). Daily linear predictability is modest, suggesting that most structure emerges over longer horizons. Spectral analysis of $x_t$ reveals statistically significant weekly and annual cycles, together with a small number of peaks at periods of approximately 3.5, 824, 929, and 1,054 days. We interpret these regular oscillations as reflecting the news production schedule and remove them before the scaling analysis by estimating and subtracting a seasonal component via standard methods (see Methods). 

The distribution of sentiment scores is broad and non-Gaussian. Kolmogorov-Smirnov (KS) and Shapiro-Wilk (SW) normality tests reject Gaussian marginals for both series. Furthermore, Hill tail exponents on the upper tails of absolute standardized residuals give $\hat{\alpha}_x \approx 4.28$ and $\hat{\alpha}_{\Delta x} \approx 3.76$, implying finite variance in both cases, with moderately heavier tails for daily changes than for the index itself. We apply a standard Hampel filter before the scaling analysis to limit the influence of isolated extreme observations (see Methods).

\subsection*{Slow–moving sentiment and short–lived daily swings}

\begin{figure}
\centering
\includegraphics[width=\columnwidth]{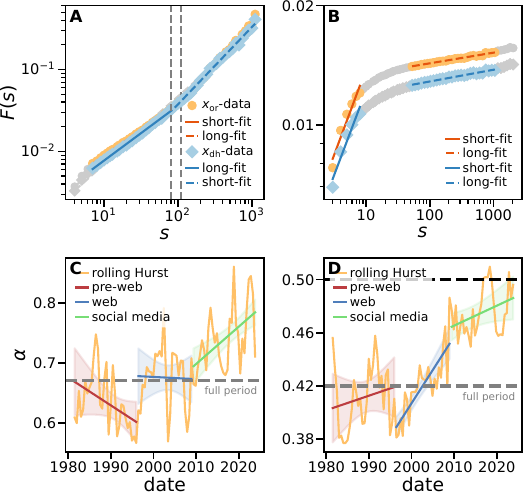}
\caption{Estimated Hurst exponents. Log–log fluctuation functions for (\textbf{A}) the raw sentiment index $x_t$ (circles) and the deseasonalised, Hampel-filtered series $x_t^{\text{dh}}$ (diamonds), with vertical dashed lines for the automatically detected crossover scales, and (\textbf{B}) the first differences $\Delta x_t$ (circles) and $\Delta x_t^{\text{dh}}$ (diamonds), with two fixed scale intervals (2–8 and 49–1095 days). Grey points were excluded from fitting. Rolling Hurst exponents for (\textbf{C}) the index at short scales (7–$s^\ast$ days) and (\textbf{D}) first differences at short scales (2–8 days). Orange lines are the empirical Hurst estimates. Shaded regions indicate 95\% Newey-West confidence intervals for the linear trends, adjusted for overlapping windows. Dashed lines mark the full sample $\alpha=H$.}
\label{fig:02}
\end{figure}

To quantify memory across timescales, we estimate Hurst exponents using detrended fluctuation analysis (DFA), which measures how fluctuations grow with the observation scale and therefore how strongly sentiment depends on its past (see Methods). DFA on the raw sentiment index suggests two–segment scaling with a crossover around $s \approx 110$ days but an unrealistically low large–scale exponent ($H_x \approx 0$), which would imply strong anti–persistence in the slow component of sentiment (Fig.~\ref{fig:02}A). This estimate is inconsistent with the smooth low–frequency evolution and likely reflects residual publication cycles and extreme days. This motivates using the preprocessed series when characterizing long-range persistence. The deseasonalised, Hampel-filtered series $x_t^{\text{dh}}$ shows a crossover at $s \approx 81$ days (Fig.~\ref{fig:02}A). The short–scale regime ($7 \le s \le 81$ days) has $H_x \approx 0.67$, indicating weak persistence, i.e.\ positive (negative) sentiment tends to be followed by similar tone over weekly–to–quarter intervals. At longer scales ($81 < s \le 1095$ days), the exponent increases towards unity $H_x \approx 0.95$, approaching the $1/f$ boundary of long-memory dynamics. In this regime, fluctuations in the sentiment index decay more slowly than exponentially and remain persistent  over multi–month to multi–year horizons.

The fluctuation curve for $\Delta x_t$ bends sharply after about a week (Fig.~\ref{fig:02}B), so we analyze one fixed interval for weekly behavior (2–8 days), and another for longer-horizon variation in daily shocks (49–1095 days). In the weekly band, increments are mildly anti–persistent with $H_{\Delta x} \approx 0.43$ for the raw differences and $H \approx 0.42$ for the pre–processed differences $\Delta x_t^{\text{dh}}$, indicating that positive shocks tend to be followed by partial reversals rather than like–signed moves. At longer horizons, the differences are nearly memoryless ($H_{\Delta x} \approx 0.03$).

\begin{table}[tp]
\centering
\scriptsize
\setlength{\tabcolsep}{6pt}
\renewcommand{\arraystretch}{1.2}

\begin{tabular}{l r@{\,}l r@{\,}l}
\hline
\multirow{2}{*}{\textbf{Model}} &
  \multicolumn{4}{c}{\textbf{Index - DFA2}} \\
\cline{2-5}
 & \multicolumn{2}{c}{\textbf{7-81 days}}
 & \multicolumn{2}{c}{\textbf{81-1095 days}} \\
\hline
empirical
  & $0.67$ & [0.66,\,0.68]
  & $0.94$ & [0.92,\,0.97] \\
shuffle
  & $0.53$ & [0.52,\,0.55]$^{\ddagger}$
  & $0.50$ & [0.45,\,0.55]$^{\ddagger}$ \\
block
  & $0.68$ & [0.66,\,0.71]$^{\times}$
  & $0.15$ & [$-0.33$,\,0.62]$^{\times}$ \\
AR(1)
  & $0.73$ & [0.72,\,0.75]$^{\ddagger}$
  & $0.52$ & [0.47,\,0.57]$^{\ddagger}$ \\
IAAFT
  & $0.67$ & [0.67,\,0.67]$^{\times}$
  & $0.97$ & [0.95,\,0.98]$^{\dagger}$ \\
\hline
\multirow{2}{*}{\textbf{Model}} &
  \multicolumn{4}{c}{\textbf{First differences - DFA1}} \\
\cline{2-5}
 & \multicolumn{2}{c}{\textbf{2-8 days}}
 & \multicolumn{2}{c}{\textbf{49-1095 days}} \\
\hline
empirical
  & $0.43$ & [0.28,\,0.58]
  & $0.03$ & [0.03,\,0.03] \\
shuffle
  & $0.40$ & [0.39,\,0.41]$^{\ddagger}$
  & $0.005$ & [0.005,\,0.006]$^{\ddagger}$ \\
block
  & $0.43$ & [0.41,\,0.44]$^{\times}$
  & $0.05$ & [0.04,\,0.06]$^{\ddagger}$ \\
AR(1)
  & $0.57$ & [0.56,\,0.59]$^{\ddagger}$
  & $0.01$ & [0.01,\,0.01]$^{\ddagger}$ \\
IAAFT
  & $0.43$ & [0.42,\,0.43]$^{\times}$
  & $0.03$ & [0.03,\,0.04]$^{\ddagger}$ \\
\hline
\end{tabular}
\caption{Hurst exponents estimated on daily news sentiment. Each cell reports $H$ with 95\% confidence intervals. Empirical intervals are $95\%$ regression C.I.s from DFA slope fits. Null model intervals are percentile intervals from 500 realizations. Superscripts denote two-sided Monte Carlo $p$-values: $^{\times}$ for $p>0.05$, $^{\dagger}$ $p<0.05$, $^{\ddagger}$ $p<0.01$.}
\label{tab:02}
\end{table}

We benchmark the results against four null models preserving different features of the time series (see Methods). Table~\ref{tab:02} shows that in the short-scale regime, the exponent $H_x=0.67$ differs significantly from i.i.d.\ shuffle and parametric AR(1) nulls (both $p<0.01$), but is statistically indistinguishable from a moving–block bootstrap and IAAFT models preserving the empirical spectrum and marginal distribution. Thus most of the weak persistence at weekly–to–quarter scales can be explained by short–range dependence and the observed power spectrum, with little evidence for additional non-linear structure. In the long-scale regime, the exponent $H_x=0.94$ is well above the i.i.d.\ and AR(1) nulls (both $p<0.01$), showing that the second scaling interval is not explained simply by independent fluctuations or short-memory autoregression. However, the estimate is close to the spectrum-preserving IAAFT model. The small difference between empirical and IAAFT exponents (on the order of $0.03$) suggests that most of the long-scale exponent is already captured by the empirical low-frequency spectrum, rather than by higher-order nonlinear structure. The block bootstrap, which reorders annual segments without preserving the full spectrum, generates much smaller exponents ($H_x\approx0.15$), because it disrupts the global low-frequency organization of the series. Thus, the long-scale regime should be interpreted mainly as strong linear low-frequency organization in the sentiment index, with only modest residual differences from a spectrum-preserving baseline. For the first differences, mild anti-persistence at weekly scales and very weak scaling at longer horizons are fully compatible with short-range dependence and the empirical spectrum. That is, while the sentiment index exhibits slow drift, the high-frequency shocks remain close to short-memory noise.

For self-affine processes with a single scaling regime, the exponents are related by $H_{\Delta x} \approx 2H_x - 1$. Our estimates show that this mapping is only locally valid, because the index series exhibits a clear crossover. At weekly–to–quarter scales, $H_x \approx 0.67$ while $H_{\Delta x} \approx 0.42$, capturing a mildly persistent macroscopic state with anti-persistent, self-correcting day-to-day shocks. Beyond roughly 3 months (the crossover), the mapping no longer holds, since the signal enters a second regime ($H_x \approx 0.94$, close to the $1/f$ boundary), whereas increments are nearly memoryless ($H_{\Delta x} \approx 0.03$).

These findings indicate a scale-dependent integrated signal in which daily U.S. news sentiment carries slowly decaying memory, while daily increments behave as short-lived perturbations. Up to roughly three months, the background sentiment shows weak long-range dependence, consistent with the persistence of slowly evolving narrative frames. At longer scales, imbalances in positive versus negative coverage accumulate into a slowly drifting state, while the increments capture high-frequency perturbations that tend to self-correct rather than compound. Sustained changes arise when many small directional nudges align over extended periods, rather than from isolated news shocks.

\subsection*{Drifting persistence in news sentiment across media eras}

To examine whether persistence changes over time, we compute Hurst exponents in rolling windows of 1,095 days (shifted by 182 days) and track their evolution. We then regress these local exponents on calendar time, splitting into three historical periods: 1980–1995 (pre-www), 1996–2008 (www), and 2009–2025 (social media platforms). These reference periods correspond to broad changes in news production and distribution. The pre-web period corresponds to a mainly print-centered news environment, the web period to the expansion of online publication and faster update cycles, and the social-media period to platform-mediated circulation, audience metrics, and stronger feedback between publication and engagement.

At short scales ($7\le s\le s^\ast$), the rolling exponents for the index display a pronounced U-shaped evolution (Fig.~\ref{fig:02}C). Persistence weakens in the pre-web period ($\beta=-0.0046$ per year, 95\% CI $[-0.0085,-0.0008]$), is essentially flat during the early web years ($\beta=-0.0004$, $[-0.0060,0.0053]$), and increases again after 2009, with a positive and significant trend ($\beta=0.0062$, $[0.0021,0.0103]$). From the fitted linear trends, the start-to-end fitted changes in the Hurst exponent are $-0.067$ Hurst units in the pre-web period, $-0.005$ in the web period, and $+0.090$ after 2009. At longer scales ($s^\ast\le s\le 546$), trends are weak and not statistically different from zero, indicating that very slow organization remains broadly stable (see SI). The crossover breakpoint $s^\ast$ typically lies between about 3 and 8 months (median $145$ days, IQR: $[96.5$,$233.5]$ days).

Trends are more pronounced for the first differences (Fig.~\ref{fig:02}D). The series remains anti-persistent ($H<0.5$) at weekly scales, but the short-scale slopes are indistinguishable from zero pre-web ($\beta=0.0011$, 95\% CI $[-0.0010,0.0031]$), clearly positive in the web period ($\beta=0.0050$, $[0.0035,0.0066]$), and smaller but still weakly positive after 2009 ($\beta=0.0015$, $[0.0000,0.0029]$). The start-to-end linear fitted changes in the Hurst exponent are positive in all periods: $\approx 0.016$ (pre-web), $0.063$ (web), and $0.022$ (after 2009). Because $H<0.5$ at the weekly scale, daily shocks still self-correct but the upward trends indicate weaker mean reversion over time. Positive slopes are also observed at longer scales, being strongest in the pre-web and social-media periods (see SI).

The rolling analysis therefore indicates that the main empirical change is not a drift in average sentiment, but a change in the response time of the sentiment process. Short-scale persistence rises after a period of pre-web weakening, while the very slow organization of the sentiment index remains relatively stable. Daily shocks still dissipate rather than build into permanent trends, but they fade more slowly, especially at weekly horizons, leaving a longer residual imprint before disappearing. As the Hurst exponents increase in the social-media period, stretches of positive or negative sentiment become easier to sustain, suggesting a shift from a more reactive news-sentiment process toward a more state-dependent one, in which current tone increasingly conditions future tone over weekly-to-quarterly horizons. 

\subsection*{A mesoscopic endogenous-memory model of news sentiment}

The estimated Hurst exponents suggest that daily news sentiment behaves like a coarse-grained observable of underlying sentiment dynamics driven by high-frequency shocks. We capture this behavior by proposing a minimal mechanistic model in which sentiment is the superposition of a slow latent sentiment component $x^L_t$ and a fast shock $y_t$ at time $t$:
\begin{equation}
x_t = x^L_t + y_t
\label{eq:02}
\end{equation}

This decomposition separates slow changes in background sentiment from short-lived daily fluctuations. The model is intended as a descriptive representation of the observed temporal structure. The slow component $x^L_t$ captures a slowly evolving sentiment background or macro-narrative component. We model it as a fractional Gaussian noise process, a standard representation of long-memory dynamics, with period-specific memory parameter $d_k$ in period $k$. The corresponding Hurst exponent is thus $H_k = 0.5 + d_k$ and the autocovariance decays as a power law in the day-lag. Larger values of $d_k$ correspond to a heavier-tailed memory profile and a more persistent carryover of past sentiment into future coverage.

\begin{figure}
\centering
\includegraphics[width=\columnwidth]{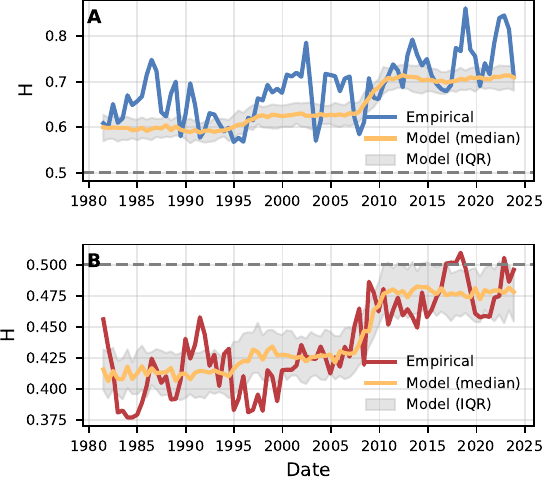}
\caption{Empirical and modeled rolling Hurst exponents. Panels show the rolling (\textbf{A}) DFA2 slopes for the sentiment index on scales 7–81 days and (\textbf{B}) DFA1 slopes for first differences on scales 2–8 days. We show the median and interquartile range (shaded area) from 100 random realizations.}
\label{fig:03}
\end{figure}

The fast component $y_t$ captures high-frequency shocks and local corrections. We model $y_t$ as a driftless AR(1) process with a period-specific feedback coefficient $\phi_k$ and a weak ($\omega>0$) dependence of shock ($\varepsilon_t$) amplitude on the previous state:

\begin{equation}
y_t = \phi_k\,y_{t-1} + \sqrt{1 + \omega y_{t-1}^2}\,\varepsilon_t,
\qquad \varepsilon_t \sim \mathcal{N}(0,1)
\label{eq:03}
\end{equation}

The parameter $\phi_k$ represents the short-range feedback gain. A more negative $\phi_k$ generates anti-persistent daily increments and local mean reversion, while more positive $\phi_k$ weakens this corrective feedback and allows shocks to relax more slowly. When $|\phi_k|$ is close to zero, shocks die out faster.

Figure~\ref{fig:03} shows that the model, calibrated using an optimisation procedure, reproduces the main temporal organization of the DFA exponents at shorter scales. For the sentiment index, empirical and model exponents on 7–81 day scales show good agreement (median correlation $\approx$$0.56 [0.48,0.60]$, RMSE $\approx$$0.069 [0.065,0.072]$, median negative bias $\approx$$-0.033 [-0.038,-0.028]$, where brackets denote the interquartile range Q1–Q3). Agreement is stronger for first differences on 2–8 day scales (respectively $\approx$$0.59 [0.54,0.65]$, $\approx$$0.035 [0.033,0.037]$, and $\approx$$0.006 [0.003,0.008]$). The best-fitting parameter set achieves a median global loss $\mathcal{L} \approx 1.54$ (see Methods). The model slightly underestimates short-scale persistence in the early periods and slightly overestimates it for short-scale differences in the most recent period, but it captures the overall drift and level of persistence across time. The calibrated values show that the model can reproduce the observed scaling patterns, but they should not be seen as unique estimates of the true underlying process.

The endogenous-memory mechanism implies that sentiment memory strengthens across periods ($d_k \approx (0.28,0.36,0.45)$), producing a longer-lived slow component. At the same time, short-run corrective feedback weakens, with $\phi_k$ moving from weak local correction ($\phi_1 \approx -0.03$) in the pre-web period to nearly white-noise dynamics ($\phi_2 \approx 0.04$) and then short-run persistence ($\phi_3 \approx 0.26$) in the social media period. Past sentiment is therefore integrated over longer horizons, while daily shocks are less strongly counteracted and can eventually be reinforced. These two interpretable mechanisms are sufficient to reproduce the observed shift from event-reactive (early period) to state-dependent sentiment dynamics (recent times), i.e.\ longer-lived sentiment in $x_t$ and weaker short-run correction in $\Delta x_t$. The stochastic model should therefore be interpreted as a minimal generative mechanism, not as a uniquely identified structural model because additional mechanisms may also contribute to the dynamics.

\subsection*{Bimodality, persistence, and the dynamics of news sentiment}

\begin{figure}
\centering
\includegraphics[width=\columnwidth]{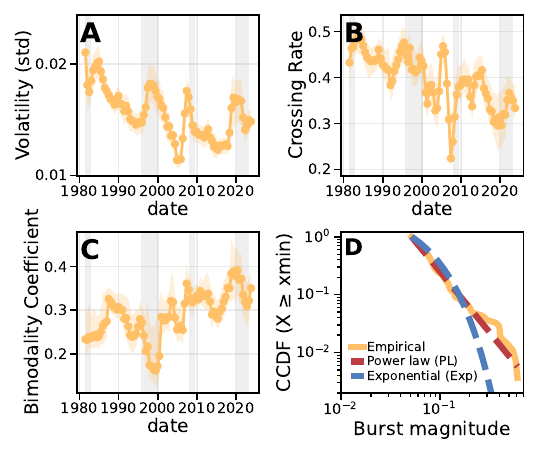}
\caption{Rolling organization and burst statistics of daily news sentiment. Panels show 1,095-day rolling within-window measures (shifted 182 days) of (\textbf{A}) volatility (within-window standard deviation), (\textbf{B}) zero-crossing rate, and (\textbf{C}) bimodality coefficient, with 95\% C.I.. Gray shading marks major macroeconomic events (Volcker disinflation, dot-com bubble, global financial crisis, COVID-19). (\textbf{D}) The complementary cumulative distributions (CCDFs) for negative sentiment bursts of burst magnitudes (sum of the absolute daily sentiment within a burst). Axes are in log scale.}
\label{fig:04}
\end{figure}

The scaling analysis characterizes how fluctuations aggregate across time, i.e.\ signal persistence across multiple scales. To contrast this with more local organization, we compute three rolling metrics that capture different within-window aspects of the dynamics (Fig.~\ref{fig:04}). Volatility, defined as the within-window standard deviation of sentiment, captures amplitude variations. It declines from roughly $0.020$ to $0.014$, implying that typical day-to-day moves have become smaller. The zero-crossing rate, which measures how often (demeaned) sentiment changes sign, falls from about $0.47$ to $0.35$, indicating fewer sign flips and longer runs of like-signed days. The bimodality coefficient, $\mathrm{BC} = (\nu^2 + 1)/\kappa$ (with index skewness $\nu$ and kurtosis $\kappa$), captures the shape of the distribution of sentiment scores. It rises from nearly $0.24$ to $0.35$, indicating that sentiment values concentrate more often in clearly positive or negative ranges and less around neutral levels. These shifts indicate that daily sentiment has become less volatile but more organized into extended positive or negative phases. These patterns are consistent with the scaling results. As short-scale persistence in the sentiment level strengthens, the within-window statistics indicate that news sentiment spends more time in positive or negative configurations and less time fluctuating around a single center.

\subsection*{Heavy-tailed, asymmetric burst dynamics of news sentiment}

Daily news sentiment has intermittent bursts of unusually high or low values. We define positive and negative bursts separately, where $x_t$ exceeds a high quantile ($q>0.9$) or falls below a low quantile ($q<0.1$). The inter-burst interval $\Delta T_i$ is the elapsed time between bursts. The burst-start process is overdispersed, with burstiness $(\sigma_{\Delta T} - \mu_{\Delta T})/(\sigma_{\Delta T} + \mu_{\Delta T}) \approx 0.35$ indicating more variability than a Poisson process, and positive serial dependence (lag-1 memory $\mathrm{corr}(\Delta T_i,\Delta T_{i+1}) \approx 0.25$), i.e.\ long and short gaps tend to follow gaps of similar length and bursts tend to arrive in temporally clustered episodes rather than independently over time~\citep{KarsaiEtAl2018}.

The variability of burst counts grows faster than their mean when we move from monthly to quarterly windows. Inter-event times $\Delta T_i$ and burst magnitudes (sum of within-burst $|x_t|$) are strongly heavy-tailed, with power-law exponents $\gamma \approx 2.5$–$4.0$ (Fig.~\ref{fig:04}D), and steeper tails for positive sentiment. Burst durations (consecutive days in a burst) show sign-asymmetry with positive bursts well described by a geometric function (short, near-memoryless durations). In contrast, negative durations have heavier tails (longer-lasting episodes) and are less cleanly separated between power-law and geometric fits (see SI).

Because the inter-event exponents satisfy $\gamma>2$, heavy-tailed renewal statistics alone are unlikely to generate the observed large Hurst exponents. Persistent sentiment regimes and sign-specific burst dynamics must also contribute. The heavy tails of burst magnitudes show that bursts, especially negative ones, can be large, while shorter, near-geometric positive durations help reconcile strong long-run persistence in the index with anti-persistent daily increments. This asymmetric avalanche structure (large, heavy-tailed negative bursts and shorter but frequent positive episodes) allows strongly signed sentiment regimes to occupy a larger fraction of time, making it easier to sustain extended periods of positive or negative coverage, consistent with longer residence times in positive or negative sentiment states.

\section*{Conclusions}

We analyzed an economic and financial news-based sentiment index of U.S.\ newspapers over 45 years. Our results show that the temporal memory of this sentiment index has lengthened over the past four decades. Daily shocks have become smaller but more temporally organized, short-scale persistence has strengthened, and aggregate sentiment now exhibits longer residence times in optimistic and pessimistic regimes. In our model, this shift corresponds to stronger endogenous sentiment memory and weaker short-range corrective feedback, a combination that produces more sustained sentiment regimes without changes in mean tone. These results identify a change in the temporal structure of the sentiment index, albeit not the exact causal mechanisms producing that change.  In dynamical terms, when memory deepens and short-range corrective feedback weakens, fluctuations can become more self-sustaining and harder to reverse~\citep{Sasahara2020, NotarmuziEtAl2022}.

The sentiment index excludes broadcast, non-U.S.\ sources, and social media and therefore does not capture the full global information ecosystem. The sentiment series relies on lexicon-based scoring to ensure temporal comparability across four decades of text. Although large language models may improve contextual classification in contemporary corpora, their training data and evolving architectures complicate consistent historical measurement. Although the original sentiment model was evaluated against manually rated articles from the historical corpus, this does not eliminate the possibility of semantic drift, changes in idiomatic usage, or in the intensity of economic language over time. Lexicon methods provide a transparent and stable benchmark for long-run structural analysis of the sentiment index using detrended fluctuation analysis. The newspaper $\times$ article-type fixed effects improve comparability by removing persistent baseline differences across outlets and article types, but they do not control for changes in article volume, sourcing, topic composition, genre mix, or time-varying editorial practices. Because we use the aggregate daily index, these remaining composition effects cannot be separated directly from changes in sentiment dynamics.

The media-period stratification should be seen as a descriptive periodization rather than as causal identification. Changes in online publication and platform distribution, audience feedback, and the persistence of narratives may all contribute to the observed persistence shift. Major events provide useful context, but separating macroeconomic persistence from media-specific dynamics would require a separate benchmarking analysis. Regular publication cycles are removed from the series before the scaling analysis through spectral methods, but other editorial and publishing meso-structures, such as article selection and repeated coverage of ongoing stories, are difficult to remove with the aggregate index and may also contribute to the persistence of sentiment. The proposed endogenous-memory model generates synthetic sentiment time series from two interpretable mechanisms, slow memory and short-run corrective feedback. The resulting series reproduce the empirical trends in the Hurst exponents of both the index and its first differences. The model is simple and not uniquely identified. A more detailed model could include non-linear amplification from external factors such as macroeconomic dynamics or (online) audience feedback. The null models provided simpler baselines for the scaling results by testing whether the observed exponents can be explained by independence, short-memory AR(1) dependence, annual block structure, or the empirical spectrum alone. More advanced statistical models, such as ARFIMA, could include extra parameters to provide better fits to the autocorrelation structure or the spectrum but with less mechanistic interpretation.

These findings matter for any model that uses news sentiment as an input. If sentiment has become more persistent, the current value of a sentiment index cannot be interpreted only as an instantaneous reaction to events or used as a predictor based only on its current level. It also reflects the accumulated memory of non-recent shocks, depending on how slowly the sentiment process relaxes. Increasing persistence changes the effective time scale of news sentiment. Even when daily changes are small, repeated weak perturbations can accumulate into longer sentiment regimes, i.e.\ amplify small signals into persistent states. Forecasting and expectation models, and risk-monitoring systems, for example in finance, macroeconomic nowcasting, and consumer-confidence monitoring, should therefore account for the response time of the sentiment process, not only its average level or daily variations.

\section*{Methods}

\subsection*{Time series pre-processing}

To isolate persistent temporal structure, we process the sentiment series as follows. Periodicities at weekly, annual, and nearby harmonics are detected from the spectrum of $x_t$ and removed via harmonic regression, creating a deseasonalized series $x_t^{\text{d}}$ (see SI). To limit the influence of rare extreme days while leaving typical fluctuations unchanged, we apply a Hampel filter to $x_t^{\text{d}}$ with a $\pm7$-day window and a 3-$\sigma$ threshold, flagging $\sim 2.5\%$ of days and replacing those values with the local median. The deseasonalized, Hampel-filtered series $x_t^{\text{dh}}$ is used for the scaling and model analyses, and the raw index $x_t$ for all other statistics. Sensitivity checks show that the DFA estimates are stable across alternative Hampel-filter windows and thresholds (see SI).

\subsection*{Detrended fluctuation analysis}

To quantify persistence and long-range dependence, we use detrended fluctuation analysis (DFA) \citep{PengEtAl1995,Kristoufek2015}. For a time series $\{x_t\}_{t=1}^N$ we construct the profile:
\begin{equation}
Y(k) = \sum_{t=1}^k (x_t - \bar{x}), \qquad k = 1,\dots,N
\end{equation}
We then partition $Y$ into non-overlapping windows of length $s$, detrend each window with a polynomial of order $m$, and compute the root-mean-square fluctuation $F(s)$ of the residuals. If fluctuations are scale-free, $\log F(s) = \alpha \log s + c$. For stationary, fractional-Gaussian-noise–like signals, $\alpha = H \in (0,1)$. For nonstationary, fractional-Brownian-motion–like signals with $\alpha > 1$, the implied Hurst exponent of the increments is $H = \alpha - 1$. We report $\alpha$ for rolling windows and interpret $H$ as appropriate in the main text.

We apply DFA of order $m=2$ (DFA2) to the sentiment index $x_t$ to attenuate slow drifts and low-frequency curvature. Fluctuation functions $F(s)$ are evaluated on logarithmically spaced scales from $s_{\min} = 7$ up to $s_{\max} \le 1,095$ days (three years), with at least four windows per scale. We allow at most one crossover. Candidate breakpoints $s^\ast$ in the range $7 \le s^\ast \le 546$ days are considered on the log-scale of $s$. For each $s^\ast$, we fit single- and two-slope models in $\log F(s)$ vs.\ $\log s$ and select the preferred specification using corrected AIC and a nested $F$ test. When a two-slope model is chosen, we retain the exponents on 7–$s^\ast$ and $s^\ast$–1095 days (SI).

For the first differences ($\Delta x_t$), we define a new profile (replacing $x_t$ by $\Delta x_t$). We use DFA of order $m=1$ (DFA1) to isolate short-run shock structure. The fluctuation curves bend near weekly scales and thus we predefine two intervals, 2–8 days and 49–1,095 days, and fit a single slope in each interval. Because $\Delta x_t$ is stationary, we interpret these slopes directly as Hurst exponents.

For the rolling analysis, we apply the same DFA procedure to $x_t$ within 1,095-day windows shifted by 182 days, restricting the maximum scale to 546 days to ensure at least four segments per scale (see SI). We estimate trends in the rolling-window Hurst exponents using OLS. Trend C.I. and $p$-values are calculated using Newey-West standard errors with lag 6 ($\approx 1095/182$) to account for overlapping windows.

\subsection*{Null models}

We compare the empirical Hurst exponents to four null models. The baseline shuffle (i.i.d.) model randomly permutes $x_t^{\text{dh}}$, preserving the marginal distribution but destroying temporal dependence. The moving-block bootstrap (block) model concatenates 364-day (52-week) blocks sampled with replacement, preserving local dependence and weekly patterns while disrupting longer-range order. The AR(1) model simulates an autoregressive process fitted by ordinary least squares to $x_t^{\text{dh}}$, preserving short-range linear dependence. The IAAFT model uses the iterative amplitude–adjusted Fourier transform procedure \citep{SchreiberSchmitz1996} to preserve both the empirical marginal (via rank matching) and the power spectrum (via phase randomization), implementing a linear null with the same second-order structure as the data.

We generate 500 random time series using each model, apply the same preprocessing and DFA pipeline as for the empirical data, and obtain distributions of $H$ for the four scale intervals used in the main text. Monte Carlo $p$-values and Benjamini–Hochberg false-discovery-rate control at $q=0.05$ are used for inference.

\subsection*{Mechanistic model calibration}

We calibrate $(d_1,d_2,d_3,\phi_1,\phi_2,\phi_3)$ using simulations. Parameter values are drawn using the Latin hypercube method over progressively narrower pre-specified ranges, starting at $\pm2$ standard deviations from the average empirical value for a given period. For each parameter set, we generate 30 synthetic sentiment series from the model and apply the same rolling DFA2 procedure used for the empirical series for levels (7–81 days) and DFA1 for first differences (2–8 days). We select the parameter set that best reproduces the empirical rolling Hurst exponents of both the sentiment index and its first differences by minimizing a global dimensionless loss function using $\mathcal{L} = \frac{|\mathrm{bias}_x|}{\sigma_x} + \frac{|\mathrm{bias}_{\Delta x}|}{\sigma_{\Delta x}} + \bigl(1-\mathrm{corr}_x\bigr) + \bigl(1-\mathrm{corr}_{\Delta x}\bigr)$, where $\mathrm{bias}_x$ and $\mathrm{bias}_{\Delta x}$ are the mean (model–empirical) differences of the rolling DFA exponents and $\sigma_x$ and $\sigma_{\Delta x}$ are their empirical standard deviations, and $\mathrm{corr}_x$, $\mathrm{corr}_{\Delta x}$ are the corresponding Pearson correlations across windows. For the selected parameter set, we generate 100 independent realizations to estimate the variability of the rolling Hurst exponents and goodness-of-fit statistics.

\section*{Acknowledgments}
The author thanks Tobin Graf and co-authors for sharing the data and Luana de Freitas Nascimento for comments.

\bibliographystyle{unsrtnat}
\bibliography{NewsSentimentPersistence}

\end{document}